\documentstyle[epsf]{elsart}
                 
\newcommand{\lessim}{\raisebox{-0.8mm}
{\hspace{1mm}$\stackrel{<}{\sim}$\hspace{1mm}}}
\begin{document}
\begin{frontmatter}
\title{
Quark distribution 
in the pion within the
instanton liquid model}
\author[IFT,JINR]{A. E. Dorokhov,}
\author[IFT]{Lauro Tomio} 
\address[IFT]{Instituto de F\'\i sica Te\'orica, UNESP,
Rua Pamplona, 145, 01405-900, S\~ao Paulo, Brazil}
\address[JINR]{Bogoliubov Theoretical Laboratory, Joint
Institute for Nuclear Research, 141980, Dubna, Russia}
\date{\today}
\maketitle
\begin{abstract}
The leading\--\-twist valence\--\-quark distribution
function in the pion is obtained at a low normalization scale 
of an order of the
inverse average size of an instanton $\rho_c$. The momentum 
dependent quark mass and the quark\--\-pion vertex 
are constructed in the framework of the instanton liquid model,
using a gauge invariant approach.  The
parameters of instanton vacuum, the effective instanton radius and
quark mass, are related to the vacuum expectation values of the
lowest dimension quark\--\-gluon operators and to the pion low
energy observables.  An analytic expression for the quark
distribution function in the pion for a general vertex function is
derived.  The results are QCD evolved to higher
momentum\--\-transfer values, and reasonable agreement with
phenomenological analyzes of the data on parton distributions for
the pion is found.
\end{abstract}
\end{frontmatter}

\section{Introduction}

Hadron structure functions, in terms of quark and gluon distributions
specifying the fraction $xp$ of the initial hadron
momentum $p$ carried by the active parton, play an important role in
QCD inclusive processes.  Although the evolution of parton
distributions at sufficiently large virtuality $Q^{2}$ is controlled
by the renormalization scale dependence of twist-2 quark and gluon
operators within QCD perturbation theory, the derivation of the
parton distributions themselves at an initial $Q^{2}$ value from
first principles still remains a challenge.
Hence, central predictions unknown in QCD are parton distributions
at relatively low virtuality determined in a nonperturbative scheme.

There is some recent progress in the calculation of moments of the
pion and $\rho -$ meson parton distributions~\cite{PiRhoLat} within
the lattice QCD (LQCD) using Wilson fermions in the quenched
approximation, where internal quark loops are neglected.  These LQCD
predictions for the moments of the pion distribution function
confirm the results of previous analyzes~\cite{MSach88}, being also
in qualitative agreement with that extracted
phenomenologically\cite{Sutt92,ReyaPi} from experiment\cite{PiSFex}.
However, the calculated moments are still of a relatively low
accuracy.  Besides, only a few lowest\--\-order moments are
available, while the reconstruction of the $x$\--\-dependent
distributions needs, in principle, the knowledge of all moments.
Furthermore, the QCD sum rules calculations of parton distributions
in the pion are only moderately successful \cite{Bel96}, the results
being justified in a limited region of the scaling variable $x$.

Recently, the quark distribution function in the pion was obtained
\cite{DRA-KS96} in the framework of the Nambu - Jona - Lasinio (NJL)
model~\cite{NJL}.  These and similar studies are based on the
assumption that the calculation of the twist-2 matrix elements,
within the QCD inspired effective approaches, gives distributions at
a low momentum scale $\mu_0 \lessim 1$ GeV, where such effective
theories make sense.  The distributions obtained are
extrapolated to higher experimentally accessible momentum scales
using perturbative QCD, so that comparison with experimental data can be
made.
However, the problem of the NJL model is that it is
nonrenormalizable and thus, to avoid this defect, different {\it ad
hoc} assumptions about momentum cutoff parameters are introduced.

The instanton model  of the QCD vacuum (for recent review see ,
e.g., \cite{Diakonov96,Shuryak96}),  which gives the  dynamical
mechanism of chiral symmetry breaking and provides the solution of
the $U_A(1)$ problem\cite{tH}, describes  well the properties of
pion~\cite{CarlCr,DP84,DP86} and kaon~\cite{EsTam}.  Moreover,
it dynamically generates the momentum\--\-dependent effective quark
mass $M_q$ and quark\--\-pion vertex $g_{\pi qq}$, and, as a
consequence, provides inherently a natural ultraviolet cutoff
parameter in the quark loop integrals through the effective
instanton size $\rho_c$.  On these grounds, one may believe that the
instanton vacuum framework represents an important advance with
respect to NJL-type models.  The first attempt to calculate the pion
structure function within the instanton model has been made in
\cite{RysRos86}.  More recently, important progress has been
achieved \cite{DK93,DPPPW96} in calculating quark distributions in
the nucleon within instanton\--\-inspired approaches.

In the present paper, based on the quark\--\-pion dynamics in the
framework of the instanton liquid model, we calculate the
leading\--\-twist valence\--\-quark distribution in the pion at a
low normalization point of the order of the inverse average
instanton size $\rho_c$.  The calculations are performed in a
gauge\--\-invariant manner by taking into account $P-exp$ 
factor explicitly in the definition of nonlocal quantities~\cite{MHax,DEM97}
and gauging the nonlocal quark - pion interaction\cite{Tern91,Birse}.
The momentum dependent quark mass and quark - pion vertex are 
constructed. The parameters of the instanton
vacuum, effective instanton radius and quark mass, are related to
the vacuum expectation values (VEV) of the lowest dimension
quark\--\-gluon operators and to the pion low energy observables.
We derive the quark distribution in the pion and all its moments for
the general form of the effective quark - pion vertex function.  The
validity of the isospin and total momentum  parton sum rules is
ensured by the pion compositeness condition\cite{CompCond}, and it 
is consistent with the gauge invariant approach.  As the effective
instanton model is valid for values of the quark relative momentum
up to $p \sim \rho_c^{-1} \approx 0.5 - 1$ GeV, the parton
distributions calculated here are defined at this (low)
normalization point $\mu_0 \sim \rho_c^{-1}$.  The results are 
QCD evolved to higher momentum\--\-transfers, and we found reasonable
agreement with phenomenological analyzes of the data on the pion
distribution function.

The paper is organized as follows. In Sect. 2, we briefly outline
the instanton liquid model and introduce the quark\--\-pion vertex.
In Sects. 3 and 4, the parameters of the instanton vacuum model are
related to
the vacuum expectation values  of the lowest dimension
quark\--\-gluon operators and to the pion low energy observables.
Then, we derive the expressions for the moments of the pion distribution
(Sect. 5) and for the $x$-dependent distribution itself (Sect. 6),
followed by the QCD evolution to higher values of the momentum transfer.
In the last section, the results are discussed.

\section{The instanton liquid model and the quark-pion
vertex}

We start with
the nonlocal, chirally invariant Lagrangian of the instanton liquid
model\cite{Diakonov96,Shuryak96}, which describes the soft quark fields
with the soft gluon fields being
integrated out.  The corresponding action for quarks interacting
through the 't Hooft vertices\cite{tH} can be expressed in  a form similar
to that of the NJL model
\begin{eqnarray}
&&S_{inst} = \int d^4x \  \bar q(x) i\hat \partial q(x) +
\int d^4x d^4x' d^4y d^4y'\
K(x,x';y,y')\cdot \nonumber \\
&&\cdot\frac{1}{4(N^2_c-1)}\left\{ \left[\frac{2N_c-1}{2N_c}
\left(\bar q_R(x') \tau^a q_L(x)\right)
\left(\bar q_R(y') \tau^a q_L(y)\right) +
\right. \right. \label{Lint}\\ && \left. \left.
+\frac{1}{8N_c}
\left(\bar q_R(x') \tau^a\sigma_{\mu\nu} q_L(x)\right)
\left(\bar q_R(y') \tau^a\sigma_{\mu\nu} q_L(y)\right)
\right]
+ (R\leftrightarrow L) \right\} .
\nonumber\end{eqnarray}
Here, $\tau^a = (1,i\vec\tau)$ are the matrices for the flavor
space, $N_c=3$ is the number of colors, and
$$\displaystyle q_{R(L)}(x) = \frac{1\pm\gamma_5}{2}q(x)$$
are the quark fields  with definite chirality.  In the following we
neglect the current quark mass and restrict ourselves only by the
nonstrange quark sector. In
eq.~(\ref{Lint}), the kernel of the four\--\-quark interaction $
K(x,x';y,y')$ characterizes the region of the non\--\-local
quark\--\-(anti)quark instanton induced interaction. It is dominated
by the contribution of the zero mode quark wave functions in the
field of (anti-)instanton:
\begin{eqnarray}
K(x,x';y,y') &=& \int dn(\rho) d^4z dU
\left[\Big( i\hat \partial \varphi_\rho(x-z) \Big)
\Big( i\hat \partial \varphi^\dagger_\rho(x'-z)\Big)\right]\cdot
\nonumber\\
&&\cdot\left[\Big( i\hat \partial \varphi_\rho(y-z) \Big)
\Big( i\hat \partial \varphi^\dagger_\rho(y'-z)\Big)\right],
\label{Sig}\end{eqnarray}
where $\varphi_\rho(x)$ is a quark zero mode profile function.  
In eq.~(\ref{Sig}),
$n(\rho)$ denotes the density of instantons with size $\rho$, $z$ is
the position of the instanton in the configuration space, and $dU$
is the color-space phase factor.

The spin-flavor structure of the action eq.~(\ref{Lint}) is
invariant under the global axial $q(x) \to
\exp{(i\gamma_5\tau\cdot\theta)}q(x)$ and vector transformations
$q(x) \to \exp{(i\tau\cdot\theta)}q(x)$ and it anomalously violates
the $U_A(1)$ symmetry:  $q(x) \to \exp{(i\gamma_5\theta)}q(x)$.
Within the  instanton liquid model \cite{Sh8289,DP84,DP86} it is
argued that due to the long range instanton - anti\--\-instanton
interaction,   configurations with large size instantons are
strongly suppressed and the instanton density is sharply peaked at
some finite average instanton size $\rho_c$ in the form $n(\rho) =
n_c \delta(\rho - \rho_c)$.  Since the instanton liquid is assumed
to be dilute, the mean separation between instantons is much larger
than the average instanton size and the effective density $n_c$ is a
small parameter of the approach.  The values of $n_c$ and $\rho_c$
are estimated from  the phenomenology of the QCD vacuum and hadron
spectroscopy to be  $n_c \sim 1\ fm^{-4}$, and $\rho_c$ changes
within an interval $(1.5 - 2)\ GeV^{-1}$,
where  we put  less
restrictive limits on the range of values.  It is important to note
that the effective instanton size  $\rho_c$,
which defines the range of nonlocality,
serves as a natural cutoff parameter of the
effective low energy model.  Moreover, the coupling constants of the
model, eq.~(\ref{Sig}), are also expressed through the  fundamental
parameters describing the QCD instanton vacuum, $n_c$ and $\rho_c$.
The model incorporates all attractive features of the NJL model and,
at the same time, is free of arbitrariness in the choice
of the
ultraviolet cutoff procedure and physically all parameters are well
understood. These peculiarities provide important advantages
of the instanton model as compared to different versions of the NJL
model (for a review, see, {\it e.g.}, \cite{NJL}).

The instanton induced interaction of quarks is responsible for
strong spin\--\-dependent forces in hadron multiplets\cite{DKZ92}.
In particular, this force is attractive for quark\--\-antiquark
states with vacuum and pion quantum numbers, repulsive for the
singlet part of $\eta'$, and absent (in the zero mode approximation)
in the vector\--\-like channels $\rho, \omega,$ etc. If the
attraction is sufficiently large, it can rearrange the vacuum and
bind a quark and an antiquark to form a  light (Goldstone) meson
state.

To study the formation of quark\--\-antiquark bound states in the
instanton liquid, it is convenient to rewrite the four\--\-fermion
term in the action eq.~(\ref{Lint}), linearizing the bilocals $\bar
q(x) q(y)$ and $\bar q(x)\gamma_5\vec\tau q(y)$ by introducing the
auxiliary composite meson\footnote{ In this work we do not
include explicitly the diquark part of the interaction generated by
instantons.} fields $M(x)$\cite{AuxFi} (mean field approximation) by
 virtue  of the separability of the four\--\-quark kernel.  Then, we
arrive at the following form of the effective nonlocal action
corresponding to eq.~(\ref{Lint}):
\begin{equation}
S = S_0 + S_{int},
\label{S}\end{equation}
where $S_0$ is the free action for quark and meson fields
\begin{eqnarray}
S_0=\int d^4x \
\left\{\bar q(x)\ i\hat D q(x) +
\frac{1}{2}\left[\sigma(x)(\Delta - m_\sigma^2)\sigma(x)\right] +
\frac{1}{2}\left[\vec\pi(x)(\Delta -
m_\pi^2)\vec\pi(x)\right]\right\},
\label{SM}\end{eqnarray}
and $S_{int}$ is the quark\--\-meson interaction part
\begin{eqnarray}
S_{int}=&&
- \int d^4X d^4x_1 d^4x_2\ F(x_1,x_2;\mu_0^2)  \cdot \label{LqMD} \\
&&\cdot \bar q(X+x_1)\
 E(X+x_1;X-x_2)[M_q + g_{M\bar qq} (\Gamma\cdot T) M(X)]\ q(X-x_2),
 \nonumber\end{eqnarray}
with Dirac and isospin matrices for different meson states according
to
$\displaystyle  (\Gamma\cdot T)_\sigma = I\cdot I,$
$\displaystyle  (\Gamma\cdot T)_{\pi} = i\gamma_5\cdot \vec\tau$.
In eq.~(\ref{LqMD}), $g_{M\bar qq}$ is the quark\--\-meson coupling
constant and $M_q$ is the effective quark mass fixed in a gauge -
invariant manner  (see below) by the compositeness condition
eq.~(\ref{Zphi}) and  the gap equation (\ref{Gap}) in terms of
the instanton density $n_c$ and the instanton size $\rho_c$.  In
Eqs.~(\ref{SM}, \ref{LqMD}) we neglect the terms induced by tensor
interaction in eq.~(\ref{Lint}) since they do not contribute to the
scalar channels.

To ensure the gauge  invariance of the bilocal quark operators,
which enters in Eqs.~(\ref{S} - \ref{LqMD}), with respect to
external electromagnetic $A_\mu(z)$ and strong $A^a_\mu(z)$ gauge
fields, we include into eq.~(\ref{LqMD}), following  \cite{Tern91},
the path\--\-ordered Schwinger phase factors
\begin{eqnarray}
&&E(x;y) = E_\gamma(x;y) \cdot E_g(x;y), \label{Pexp}\\
&&E_\gamma(x;y) = P\ \exp\left\{-ieQ\int_x^y dz^\mu\
A_\mu(z)\right\},
\nonumber\\
&&E_g(x;y)= P\ \exp\left\{-ig\frac{\lambda^a_c}{2} \int_x^y
dz^\mu\ A^a_\mu(z)\right\}, \nonumber\end{eqnarray}
where the charge matrix is $Q=(1/3 + \tau^3_f)/2$,
and the partial derivative $\partial_\mu$ is
replaced by the covariant one $D_\mu = \partial_\mu - ieA_{\mu} - ig
A_{\mu}^a \lambda^a/2 $.  We adopt here that the integral in the
exponent is evaluated along a straight line with $P$ being the
path-ordering operator.  The incorporation of a gauge - invariant
interaction with gauge fields is of principal importance in order to
treat correctly the hadron characteristics probed by external
sources such as hadron form factors, structure functions, etc.

The Fourier transformed  gauge-invariant nonlocal
vertex function $\tilde F(k_1,k_2;\mu_0^2)$
describes  the amplitude of soft transition of a pion with momentum
$p$ into a quark  and an antiquark with momenta $k_1 = p+k/2$ and
$k_2 = p-k/2$, respectively.  This function represents the full
interaction vertex with all quark-gluon excitations, harder than the
scale $\mu_0 \sim 1/\rho_c$, strongly (exponentially) suppressed.
It is defined
\begin{equation}
F(k_1,k_2;\mu_0^2) = \sqrt{\tilde Q(k_1)\tilde Q(k_2)}.
\label{Finst}\end{equation}
through 
the quark propagator $\tilde Q(p)$ 
normalized at zero to unity 
\begin{equation}
\tilde Q(p) = \frac{1}{(2\pi)^2}\frac{p^2}{\rho_c^2}
\int d^4x\ exp{(-i p\cdot x)} Q(x^2), \;\;\; (p = |p|) \;\;\; , 
\label{Qinst}\end{equation}
where
\begin{equation}
Q(x^2) =
\langle: \bar q(0) E_g(0,x) q(x):\rangle / \langle: \bar q(0)
q(0):\rangle
\label{Qx2}\end{equation}
is the normalized instanton induced nonperturbative part of the
gauge invariant quark propagator in configuration space.
Using the explicit expressions for the instanton field and quark
zero mode the gauge-invariant quark propagator is given by
\cite{Sh8289,MHax,DEM97} 
\begin{equation}
\displaystyle Q(x^2)  = 
\frac{8\rho_c^2}{\pi} \int_0^\infty\ dr r^2 \int_{-\infty}^\infty\ dt
\frac{\cos[\frac{r}{R}(\arctan( \frac{t+|x|}{R}) - \arctan
(\frac{t}{R}))]} {[R^2+ t^2]^{3/2}[R^2+ (t+|x|)^2]^{3/2}}, \label{E24}
\end{equation}
where $R^2 = \rho_c^2 + r^2, ~r=|\vec{z}|, ~t=z_4$.
In the derivation of these equations  a reference frame is used,
where the instanton is at the origin and $x^\mu$ is parallel to one
of the coordinate axes, say $\mu = 4$, serving as a ``time" direction
$({\it i.e.},\vec x = 0,\ x_4 = |x|)$. 
The propagator has the following expansions at small and large
distances:
\begin{eqnarray}\displaystyle    
Q(x^2) = \left\{ \begin{array}{l} 
 {1 - {\displaystyle \frac{1}{4}} \frac{\displaystyle x^2}
{\displaystyle \rho_c^2} + ...\;\;\; {\rm as}
\;\;\; x^2\to
0; } 
\nonumber \\
{2\frac{\displaystyle \rho_c^2}{\displaystyle x^2}  
+ ... \;\;\; {\rm as}
\;\;\; x^2\to\infty .}
\end{array} \right.
\label{Qexpan}
\end{eqnarray} 

\begin{figure}
\vspace*{0.5cm}
\epsfxsize=15cm
\epsfysize=15cm
\centerline{\epsfbox{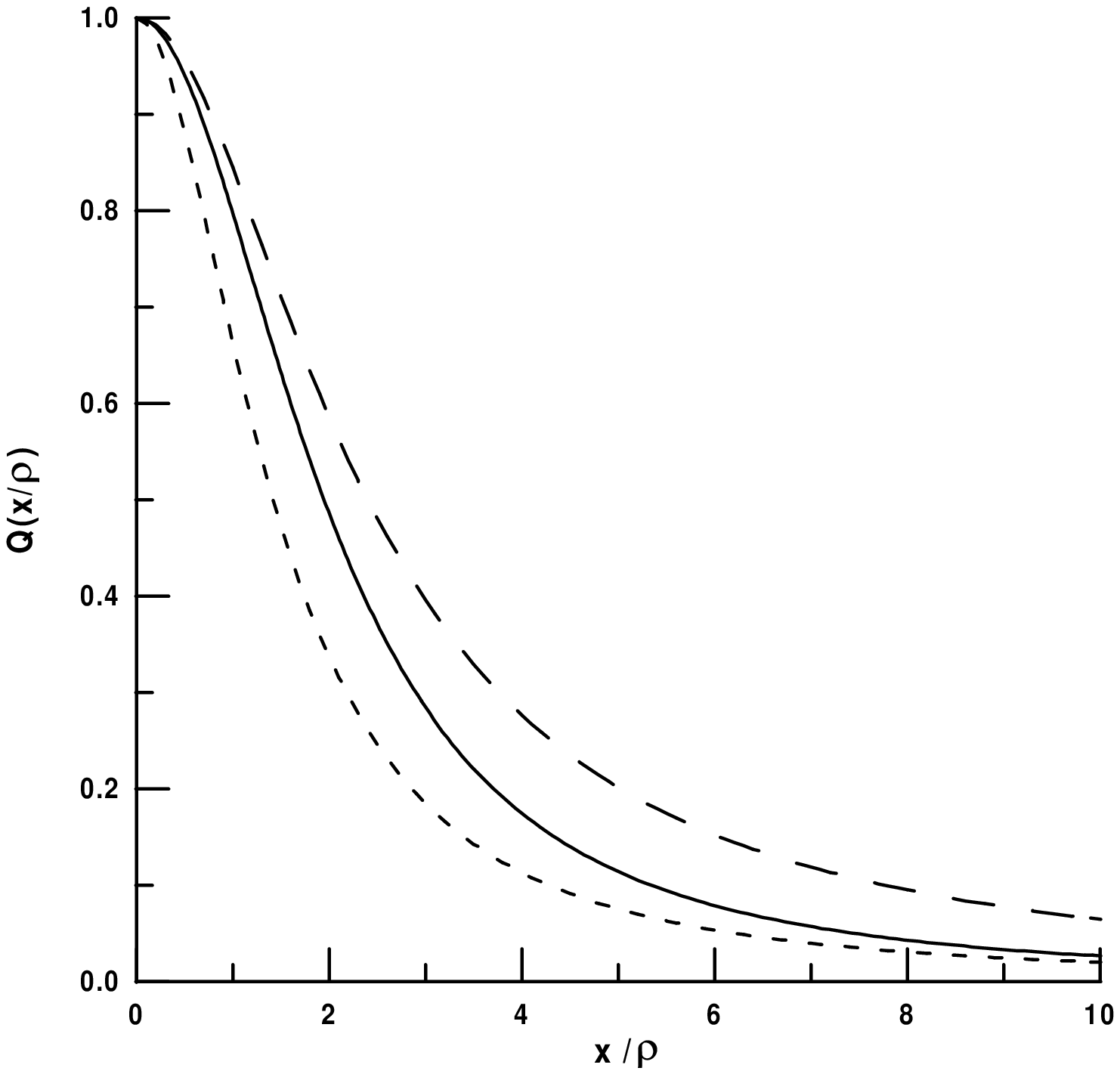}}
 \caption[dummy0]{Configuration space representation 
of the normalized instanton induced nonperturbative part of 
the gauge-invariant quark propagator, eq.(\ref{E24}) (solid line); and
the corresponding propagators derived without $P-exp$ factor in the singular
(short-dashed) and regular, eq.(\ref{Qxreg}), 
(long-dashed) gauges. 
\label{fig:Qx} }
\end{figure}

\begin{figure}
\vspace*{0.5cm}
\epsfxsize=15cm
\epsfysize=15cm
\centerline{\epsfbox{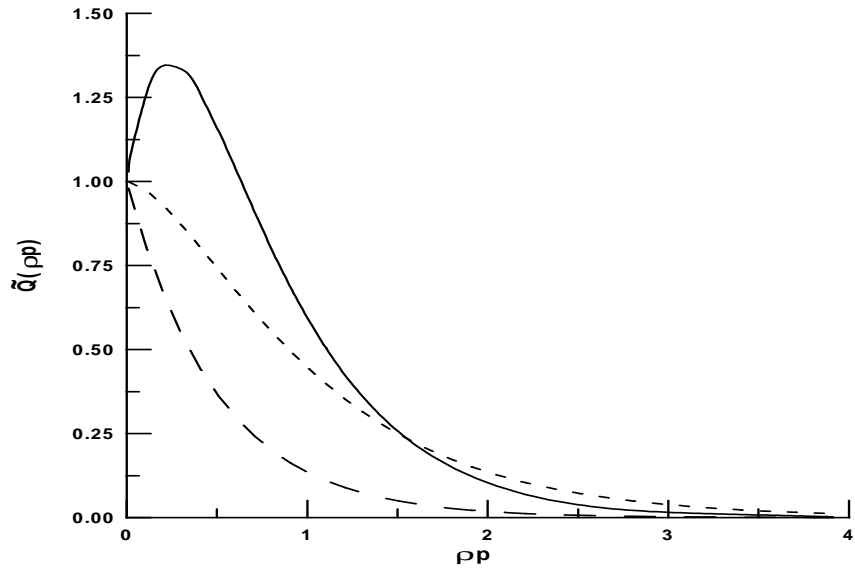}}
 \caption[dummy0]{Normalized momentum space representation for the 
same propagators given in Fig.\ref{fig:Qx}, corresponding to eqs.
(\ref{Qinst}) (solid line), (\ref{Qpreg}) (long-dashed) and
eq.(\ref{Qpsing}) (short-dashed). 
\label{fig:Qp} }
\end{figure}

The gauge-invariant quark propagators in configuration and momentum
representation
are plotted in Figs. \ref{fig:Qx} and \ref{fig:Qp}, respectively, 
along with the propagators derived in 
neglecting $P-exp$ factor in (\ref{Qx2}) and using the expressions for
the quark zero mode in the singular and regular gauges. In the regular 
gauge, one has
\begin{equation}
Q_{reg}(x^2) = \left.
\frac{2}{y^2}\left(1-\frac{1}{\sqrt{1+y^2}}\right)
\right|_{\displaystyle y = \frac{\displaystyle x}{
\displaystyle 2\rho_c}}.\label{Qxreg}
\end{equation}
In the momentum representation, the
normalized quark propagators (without $P-exp$ factor) are proportional
to the square of the quark zero mode in the corresponding regular 
and singular gauges:
\begin{eqnarray}
\tilde Q_{reg}(p) &=& \exp{(-2\rho_c p)} 
\label{Qpreg} \\
\tilde Q_{sing}(p) &=& \left\{ \left. 
z\frac{d}{dz}\left[ 
I_1(z)K_1(z) - I_0(z)K_0(z)\right]
\right|_{\displaystyle z = 
\frac{\displaystyle \rho_c p}{\displaystyle 2}}
\right\}^2 .
\label{Qpsing}
\end{eqnarray} 
From Fig. \ref{fig:Qp}, one can observe
that in the momentum representation the shape of the propagator is
very sensitive to the $P-exp$ factor~\footnote{
To avoid inconsistency with gauge invariance, one cannot use 
the treatment of the quark propagator 
in the $p-$representation, in factorizable form, as done 
in  ref.~\cite{MHax}.}.

One of the advantages of using the gauge - invariant
formalism is that the parameters of the model, such as the size of
instantons and the effective quark mass, gain physical meaning. 
As a consequence, all other physical quantities expressed through these
parameters become automatically gauge - invariant ones.  Moreover,
they could be compared with those calculated in the lattice QCD,
QCD sum rules or other QCD inspired approaches.
In contrast, when one deals with noninvariant - gauge objects there
can be chosen any convenient gauge. \ It is most correct to consider
the instanton vacuum field in the singular gauge and to construct the
effective action in this specific gauge~\cite{Diakonov96,Shuryak96}.
In the coordinate space
in the singular gauge  the instantons fall off rapidly enough to provide
small overlaps of neighbor pseudoparticles and quasiclassical
considerations are justified.  But at the end the action has to be
independent on the choice of the gauge, otherwise the form of the action
and other observables look rather awkward. This explicitly gauge invariant
form is suggested in Eqs. (\ref{Lint} - \ref{LqMD}). 
Note also that the quark propagator, eq.~(\ref{Qx2}), has a
direct physical interpretation in the heavy quark effective
theory of heavy-light mesons as it describes the propagation of
a light quark in the color field of an infinitely heavy 
quark~\cite{Sh8289}.

It is important to emphasize that the meson fields entering the
action eq.~(\ref{S}) are renormalized and the field renormalization
constants of composite mesons are set equal to zero,
\begin{equation}
Z_{M} = 1 - g^2_{M q\bar q}\left.
\frac{\partial {\Pi}_M(p^2)}{\partial p^2}\right|_{p^2 = - m^2_M} = 0,
\label{Zphi}\end{equation}
where $\Pi_M(p^2)$ is the meson field polarization operator.  This
condition \cite{CompCond} fixes the couplings of meson fields to
quarks, $g_{M q\bar q}$, (see section 4) and is a consequence of the
compositeness of hadron states manifesting themselves as  poles in
the quark-(anti)quark scattering amplitude.  As we will see below,
it is precisely this condition supplemented by the gauge -
invariance of the effective action, given by eq.~(\ref{S}), that leads to
the
correct parton isospin and momentum sum rules in the model.

\section{Dynamical quark mass and expectation values of quark-gluon
operators}

 Due to the effect of spontaneous breaking of the chiral symmetry,
the momentum dependent quark mass $M_q(k)$ is dynamically developed.
It obeys the well\--\-known gap equation\cite{Diakonov96,DP86}
\footnote{ Here and in the following, all Feynman diagrams are
calculated in the Euclidean space $(k^2 = -k_E^2)$ where the
instanton induced form factor is defined and rapidly
decreases, so that no ultraviolet divergences arise.  At the very end
we simply rotate back to the Minkowski space.  One can verify that
the numerical dependence of the results on the pion mass and the
current quark mass is negligible and can be dispensed with in the
following considerations.  }
\begin{equation}
\int \frac{d^4k}{(2\pi)^4}
\frac{M_q^2(k)}{k^2 + M_q^2(k)}=
\frac{n_c}{4N_c},
\label{Gap}\end{equation}
the solution of which has the form
\begin{equation}
M_q(k) = M_q \tilde Q(k).
\label{Mq(k)}\end{equation}
Given the dynamical mass, the values of the quark
condensate,
\begin{equation}
\langle\bar qq\rangle = -4N_c \int\limits \frac{d^4k}{(2\pi)^4}
\frac{M_q(k)}{k^2 + M_q^2(k)},
\label{QQcond}\end{equation}
and the average quark virtuality in the vacuum \cite{MihRad92},
\begin{equation}
 \lambda_q^2 \equiv
\frac{\langle:\bar q D^2q:\rangle}{\langle:\bar q q:\rangle} =
-\frac{4N_c}{\langle\bar qq\rangle} \int\limits \frac{d^4k}{(2\pi)^4}
k^2 \frac{M_q(k)}{k^2 + M_q^2(k)},
\label{QVirt}\end{equation}
can be found. \
The average quark virtuality defines the derivative of the quark
condensate and thus nonlocal property of it. One of the main
suggestion of the QCD sum rule method \cite{SVZ79} was that the
local quark and gluon condensates dominate in the light hadron
physics and introduction of higher dimensional corrections or even
nonlocal condensates themselves \cite{MihRad92} have not to change
the standard results too much. Thus, at least for consistency of
local and nonlocal QCD sum rules, the derivative (virtuality)
value has to be relatively small. \ Phenomenologically, there is rather fine
QCD sum rule analysis of this value \cite{BI82} where it has to be
defined as $\lambda_q^2 \approx 0.4 \pm 0.2$ GeV$^2$.
The LQCD calculation yields $\lambda_q^2 = 0.55 \pm 0.05\ GeV^2$
\cite{KSch87}.
Certainly, there is corrections from direct instantons to the QCD sum rule
result, but they have not to change the result drastically.
It would also be urgent if the LQCD estimation could be confirmed by new
calculations.
The ratio $\eta = M_q^2(\lambda_q)/\lambda_q^2$ characterizes the
diluteness of the instanton liquid vacuum.
The smallness of $\eta$ means that the dynamically generated
quark mass is not big enough essentially to modify
the instanton vacuum. 

For the moment, it is instructive to consider equations
Eqs.~(\ref{Gap} - \ref{QVirt}), neglecting the term $M_q^2(k)$
compared to $k^2$ in the denominator of the integrands. 
This approximation is justified in the dilute liquid regime where
$\langle k^2\rangle = \lambda^2_q >> M_q^2(\lambda_q)$.  
The observed accuracy of such procedure is better than $20 - 30 \%$ 
if $\eta < 1$. \ Then,
from Eqs.~(\ref{QQcond}) and (\ref{QVirt}), by using the explicit forms
given in Eqs.~(\ref{E24}), (\ref{Qinst}),  (\ref{Mq(k)}), we have
\begin{equation}
\langle\bar qq\rangle = -\frac{N_c M_q}{2\pi^2{\rho_c}^2}, \ \ \  \ \
\lambda_q^2 = \frac{2}{{\rho_c}^2}.   \label{qqap}
\end{equation}
The first relation, put in the form
$\displaystyle
M_q = -({2\pi^2}/{N_c})
{{\rho_c}^2  \langle\bar qq\rangle},
$
coincides with the result obtained in ref.~\cite{SVZ80},
where the effective quark mass has been defined in a system of small size
instantons interacting with long wave vacuum fields.  
The coefficient in this relation  is equal to the normalization factor  
of the momentum representation of the quark propagator. It turns out that
this factor, which is equal to $(2\pi)^2$, as seen in Eq.(\ref{Qinst}),
is the same which appears in the gauge-invariant propagator and also in 
the singular gauge propagator (without $Pexp$).

The second relation in eq.~(\ref{qqap}) has recently
been obtained in ref.\cite{DEM97}, where nonlocal properties of the
quark condensate are studied within the instanton model.  The same
result was also obtained in ref. \cite{PolW96} from direct
calculations of the local mixed quark-gluon condensate
in the framework of the Diakonov - Petrov model:
\begin{equation}
\frac{\lambda_q^2}{2} = \frac{\displaystyle
\langle:\bar q(ig
\sigma_{\mu\nu}G_{\mu\nu}^a\frac{\lambda^a}{2})q:\rangle}
{\displaystyle \langle:\bar qq:\rangle}.
\label{DP}\end{equation}
It is clear, from the expressions for the average quark virtuality,
that the range of the quark - antiquark interaction is characterized
by the effective size $\rho_c$ of the instanton fluctuations in the
QCD vacuum.
The natural gauge - invariant definition for the average quark virtuality,
eq.~(\ref{QVirt}) (and also that in eq.~(\ref{QQcond}) for the quark
condensate), with $M_q(k)$ defined in Eqs.~(\ref{Qinst}) and
(\ref{Mq(k)}), is valid only if the zero mode solution in
eq.~(\ref{Qinst}) is written in the gauge-invariant way. 
If we substituted its expression in the singular gauge (in neglecting
$Pexp$ factors) in eq.~(\ref{QVirt}), we would obtain $\lambda_q^2 = 9/(2
\rho_c^2)$, with a coefficient far from the correct one.

Inverting the relations eq.~(\ref{qqap}),  we express the parameters
of the instanton vacuum model in terms of the fundamental parameters
of QCD vacuum
${\rho_c}^2 = {2}/{\lambda_q^2},$   
$M_q = -({4\pi^2}/{N_c})({\langle\bar qq\rangle}/{\lambda_q^2}).$
If we expected ``standard" values for the quark condensate,
$\langle~\bar qq\rangle \approx -(230\ \mbox{MeV})^3$ (see, e.g.,
\cite{Shuryak96}), and for the average quark virtuality,
$\lambda_q^2 \approx 0.5\ GeV^2$ \cite{BI82,KSch87}, we would obtain
$\rho_c^{-1} \approx 0.5$ GeV and $M_q \approx 0.32$ GeV.
As we will see in the next section, the joined analysis of the
vacuum and pion properties confirms this estimation.

Within the dilute liquid approximation, 
the gap equation, (\ref{Gap}), leads to
\begin{eqnarray}
n_c = \frac{N_c M_q^2\lambda^2_q}{4\pi^2} I_n,
\;\;\;{\rm with}\;\;\; I_n = \int_0^\infty du u 
\tilde Q^2(u/\rho_c) \approx 0.61,
\label{ncap}
\end{eqnarray}
where the constant $I_n$ is independent of $\rho_c$.
There are other different useful combinations relating vacuum
parameters with each other.
For example, 
\begin{equation}
\langle\bar qq\rangle = - \frac{1}{I_n}\frac{n_c}{M_q}
= - \frac{1}{2\pi \sqrt{I_n}}\sqrt{N_c n_c \lambda_q^2}.
\label{Sh2}\end{equation}
These relations have the same parametric dependence as in the
estimations obtained in \cite{Sh8289,Shuryak96} but with different
coefficients.  The first one expresses the quark  condensate in
terms of the effective single instanton contribution times the
density of instantons.  The reason for the difference in the
coefficients is that in \cite{Sh8289}, where it looks as
$\displaystyle \langle\bar qq\rangle = - {n_c}/{M_q}$, the
expressions were obtained from the instanton formula in the gas
approximation by {\it ad hoc} replacing the current quark mass by
the effective quark mass.  In contrast, in deriving eq.~(\ref{Sh2})
this replacing procedure is fixed by the gap equation
eq.~(\ref{Gap}), 
with a definite coefficient mainly defined by the
slope of the form factor $M_q(k)$.
The second relation represents a self-consistent value
of the quark condensate in the instanton vacuum model (cf. 
\cite{Shuryak96}).
Further, since the instanton
contribution to the value of the gluon condensate is given by:
$\displaystyle \langle\frac{\alpha_s}{\pi}G^2\rangle|_{inst} =
8n_c$, it can be expressed through the quark condensate and the
average quark virtuality
\begin{equation}
\langle\frac{\alpha_s}{\pi}G^2\rangle|_{inst} =
\frac{2^5 \pi^2 I_n}{N_c}\frac{\langle\bar qq\rangle^2}{\lambda_q^2 }
\lessim 0.019\ GeV^4.
\label{Sh21}\end{equation}
The ``standard" value of the gluon condensate estimated in the
original work, in ref.~\cite{SVZ79}, was $\displaystyle
\langle({\alpha_s}/{\pi}) G^2\rangle\simeq 0.012$ GeV$^4$.  
The latest reanalysis
\cite{Broad94} of the QCD sum rules for heavy and light mesons and
also recent lattice results \cite{DiGiGl97} provide values which are
twice or even larger than the ``standard" one.

\section{Pion low energy observables}

Let us now consider the low-energy observables of the pion.  
The pion - quark coupling constant is determined by the
compositeness condition, eq.~(\ref{Zphi}), with the pion mass
operator being
\begin{equation}
\Pi_\pi(p^2) = N_c  \int \frac{d^4k}{(2\pi)^4}
\tilde F^2(k,k+p;\mu_0^2)
Tr\{\gamma_5S(k+p)\gamma_5S(k)\}, \label{MasOp}\end{equation}
where  the normalized nonlocal vertex  is given in (\ref{Finst})
and the quark Green function  is
$\displaystyle S(k) = [{M}_q\tilde Q(k) + \hat k]^{-1}$
with $\tilde Q(k)$ defined in (\ref{Qinst}).

From the definition eq.~(\ref{Zphi}) we derive the expression for
the pion - quark coupling constant $g_{\pi q\bar q}$
\begin{equation}
g_{\pi q\bar q}^2 =\frac{2\pi^2}{N_c I_{g\pi}(-m_\pi^2)}.
\label{gpi}\end{equation}
In the case of a massless pion, the integral $I_{g\pi}$
reduces to
\begin{eqnarray}
I_{g\pi}(0) = \int\limits_{0}^{\infty}
\frac{dk\ k^3 \tilde Q(k)}{D^2(k)}
\left[1 - \frac{k}{4} \frac{\tilde Q'(k)}{\tilde Q(k)}
+ \left(\frac{k}{4} \frac{\tilde Q'(k)}{\tilde Q(k)}
\right)^2 \right] ,
\label{I2mpi=0}\end{eqnarray}
where
\begin{equation}
\tilde Q'(k) = \frac{d}{dk}\tilde Q(k), \;\;\;\;
D(k) = M_q^2\tilde Q^2(k) + k^2.
\label{fD}\end{equation}
The expression for $g_{\pi q\bar q}$ given in
Eqs.~(\ref{gpi}) and (\ref{I2mpi=0}) agrees with that derived in
\cite{DP86}.

To fix the parameters in the instanton model, we consider the low
energy decay constants of the pion.  As it has recently been shown
in \cite{Birse}, in the presence of nonlocal separable interaction
the axial current  conserved in  the chiral limit can be constructed
from the action eq.~(\ref{Lint}) by using a Noether - like
method.\footnote{One of us (A.E.D.) thanks M.C. Birse for discussion
of the problem of current conservation in the nonlocal models.} The
full current is the sum of local,
\begin{equation}
j^{\mu a}_{5(loc)}(x) = \frac{1}{2}
\bar q(x) \gamma^\mu\gamma_5\tau^a q(x),
\label{j5loc}\end{equation}
and nonlocal,
\begin{eqnarray}
j^{\mu a}_{5(nl)}(x)  = &&   \int \prod_{i=1}^{4} dx_i\
K(x_1,x_2,x_3,x_4)
\{A_1(x_1,x_2,x_3,x_4)
\bar q(x_1) i\gamma_5\tau^a q(x_3) \bar q(x_2) q(x_4) +
\nonumber\\
&&+A_2(x_1,x_2,x_3,x_4)
i\varepsilon^{abc} \bar q(x_1) \tau^c q(x_3)
\bar q(x_2) i\gamma_5\tau^b q(x_4)\}
\label{j5nl}\end{eqnarray}
pieces. The coefficients $A_1$ and $A_2$ are derived in ref.~\cite{Birse}
and
in principle depend on the gauge fixing procedure.  Fortunately,
there is no path dependence for the longitudinal components of the
current, and thus, the decay constants considered below are
well-defined.

The axial and vector currents in different isospin states have a
similar structure \cite{Birse}. As a result, various Ward identities
which follow from (partial) current conservation and the
low\--\-energy theorems are satisfied. In particular, the Goldberger
- Treiman relation for the quark-pion coupling constant has the
usual form
\begin{equation}
g_{\pi qq} = \frac{{M}_q}{f_\pi}
\label{fpi}\end{equation}
and the $\pi^0 \rightarrow \gamma\gamma$ decay constant
\begin{equation}
g_{\pi\gamma\gamma}= \frac{1}{4\pi^2 f_\pi}
\label{Dec}\end{equation}
is consistent with the requirement of axial anomaly.

We fix the model parameters to give the pion weak decay constant
$f_\pi$, within 1\% of accuracy.  \
In Table 1, we present the results. For the two model
parameters, $M_q$ and $\rho_c$, we show the 
predictions for the quark-pion coupling $g_{\pi qq}$, the quark
condensate $\langle\bar qq\rangle$, the average vacuum quark
virtuality $\lambda_q^2$ and the instanton density $n_c$.

\vskip 0.5cm

\begin{center}
{\bf Table 1.} The values of the low energy vacuum and pion observables
discussed in the text.\\[0.5cm]
\begin{tabular}
{||c          |c|c|c|c|c|c||}  \hline \hline
$M_q$ & 
$\rho_c$ &
$f_\pi$ & 
$g_{\pi qq} $ & 
$|\langle\bar qq\rangle|^{1/3}$& 
$\lambda_q^2$& 
$n_c$ \\
(GeV)&(GeV$^{-1}$)&(MeV)&&(MeV)&(GeV$^2$)&(fm$^{-4}$)\\ 
\hline
0.18 & 1.0 & 92 & 3.7 & 294 & 2.2 & 1.56 \\
0.22 & 1.5 & 92 & 5.5 & 233 & 1.0 & 0.86 \\
0.23 & 1.7 & 93 & 6.2 & 215 & 0.83& 0.70 \\
0.26 & 2.0 & 91 & 7.7 & 197 & 0.63& 0.57 \\
\hline\hline
\end{tabular}
\end{center}

\vskip 0.5cm

As shown in Table 1, the values of the parameters $M_q$ and $\rho_c$ that
reproduce the lowest dimension VEV with an accuracy of an order of
$30\ \%$ are in the ``window" $M_q= (220 - 260)$ MeV, $\rho_c = (1.5
- 2.0)$ GeV$^{-1}$.  In the following we use the typical set of
parameters:
\begin{equation}
M_q= 230 MeV,\ \ \ \ \  \rho_c = 1.7\ GeV^{-1}.
\label{Pars}\end{equation}
The diluteness condition $\eta = M_q^2(\lambda_q)/\lambda_q^2 <<1$
is well satisfied within the whole ``window".

The momentum dependence of the vertices in the numerators of the
integrands (which are defining physical quantities) is 
important because it provides the ultraviolet regularization. 
Also, due to momentum dependence of the vertices, the measure 
in the integrals looks like  product of some powers of $k^2$ and 
the function $\tilde Q(k)$. 
This measure has maximum at $k^2$ of order $1/\rho_c^2$
and, at small momenta, the momentum
dependent quark mass in denominators can be substituted by an
effective constant mass parameter $m_q \approx M_q(k \sim
\rho_c^{-1})$.
With the form of momentum distribution shown in Fig. 2,
it approximately equals  to the mass at zero, $M_q(0)$.
This mass parameter $m_q$ has to be identified with
the standard constituent quark mass. Corresponding to this substitution, 
we redefine the function
$D(k)$ given in eq.~(\ref{fD}) as 
$D(k) = m_q^2 + k^2$.  The  choice of the mass parameter,
\begin{equation}
m_q = 230\ \mbox{MeV},
\label{mq}\end{equation}
well reproduces the integrals defining the VEV given by
Eqs.~(\ref{Gap} - \ref{QVirt}).  This constant - mass approximation
is often used in practice with the quark mass in the region $250 -
350$ MeV (see, f.i. \cite{DPPPW96,Ito92,Anik94}).  

The model parameters and predictions for
vacuum and pion observables are obtained within a set of
approximations.  
We are working in the chiral limit
of zero current quark mass. Further, within the zero mode
approximation small contributions coming from vector mesons are
neglected. Also only the lowest two - quark Fock intermediate state
in the pion is taken into account, which corresponds to the quenched
approximation.  We regard that all these factors can change a little
the numbers in Table 1, but the qualitative results discussed are
not greatly influent.

\section{Moments of the quark distribution function}

 The standard QCD analysis based on the Operator Product Expansion
(OPE) relates moments of parton distributions at a given scale to
the hadronic matrix elements of local twist-2 operators.  This
formalism is employed to separate the hard and soft parts of the
forward scattering amplitude.  Within the OPE,  the hard part is
calculable within perturbation theory in the form of Wilson
coefficients. The soft part is represented by a set of local
operators classified by the twist. Their matrix elements
accumulate information on the nonperturbative structure of the QCD
vacuum.

The twist-2 gauge - invariant non\--\-singlet local quark\footnote{
As in ref.~\cite{DPPPW96}, we will neglect gluon operators,
justified by the smallness of the diluteness parameter  $\eta$.}
operators with flavor $j$ are defined by
\begin{equation}
O_{\mu_1\mu_2..\mu_N}^{j} =
i^N \bar q_j\{\gamma_{\mu_1}D_{\mu_2}..D_{\mu_N}\}_S q_j,
\label{OPE}\end{equation}
where $D_\mu = \partial_\mu - ig A^a_\mu \tau_a$ is the covariant
derivative and the symbol $\{...\}_S$ means the traceless and
symmetric part of the tensor.  The matrix elements $A_N^j$ of the
local operators $O_N^j$ between pion states $|\pi(p)\rangle$ with
momentum $p$, renormalized at the normalization scale $\mu$, are
defined by
\begin{equation}
A_N^j(\mu^2) = \frac{i^N}{2}
\langle\pi(p)| \bar q_j \hat n (n_\nu D^\nu)^N q_j|\pi(p)\rangle|_\mu,
\label{An}\end{equation}
where $n_\nu$ is a light-like vector, with $n^2=0$ and $(np) =1$,
introduced to select the symmetric traceless part of the operator
$O_N^j$, eq.~(\ref{OPE}).  Let us now define  the quark distribution
for the $j-$th flavor in terms of its moments, viz.:
\begin{equation}
A_N^j(\mu^2) = \int^1_0 dx\ x^{N-1} q_j(x,\mu^2).
\label{Ani}\end{equation}
The $x-$variable is the fraction of the longitudinal pion momentum
carried by a quark in the infinite-momentum frame.  The $\mu^2$
dependence of $A^j_N$ is known exactly from the solution of the
perturbative QCD evolution equations, while the nonperturbative
dynamical model provides the initial input for this evolution. These
initial values of the moments are calculated here in the instanton
model, which specifies a low momentum transfer value related to the
scale $\mu_0^2 \propto 1/\rho_c^2$.

The $N-$th moment $A_N^{val}(\mu_0^2)$ in the instanton model, from
Eqs.~(\ref{OPE} - \ref{Ani}), can be expressed as (see Fig. 3)
\begin{eqnarray} \displaystyle
A_N^{val}&&(\mu_0^2) p_{\mu_1}p_{\mu_2}..p_{\mu_N} =
2 N_c g_{\pi q\bar q}^2 \int \frac{d^4k}{(2\pi)^4}
\label{MomIns}\cdot\\&&
\cdot \left\{\tilde F^2(k,k+p;\mu_0^2) Tr[\gamma_5 S(k+p)
\{ \gamma_{\mu_1} (k+p)_{\mu_2} .. (k+p)_{\mu_N} \}_S
S(k+p) \gamma_5 S(k)] - 
\right. \nonumber\\ && \left.
- 2 \Big[\frac{\partial \tilde {F^2}(k,k+p;\mu_0^2)}
{\partial(k+p)^2}\Big]
Tr[\gamma_5 S(k+p)
\{ (k+p)_{\mu_1} (k+p)_{\mu_2} .. (k+p)_{\mu_N} \}_S
\gamma_5 S(k)]  \right\}.
\nonumber\end{eqnarray}
Here, due to the nonlocal character of the interaction
Eqs.~(\ref{LqMD} - \ref{Pexp}), the additional term with a
derivative ensures gauge invariance of the approach and enables us
to satisfy the isospin and momentum conservation sum rules. Indeed,
taking into account the compositeness condition eq.~(\ref{Zphi}) we
get for the  first two moments
\begin{equation}
A_1^{val}(\mu_0^2) = g_{\pi q\bar q}^2 \cdot 
\left.\frac{\partial {\Pi}_\pi(p^2)}{\partial p^2}\right|_{p^2=0}
 = 1,
\;\; 
A_2^{val}(\mu_0^2) = \frac{g_{\pi q\bar q}^2}{2}\cdot 
\left.\frac{\partial {\Pi}_\pi(p^2)}{\partial p^2}\right|_{p^2=0}
= \frac{1}{2}.
\label{A1-2}\end{equation}
The results for the first two moments, eq.~(\ref{A1-2}),
manifest the (normalization) isospin and momentum sum rules
for the valence-quark distribution function as
\begin{equation}
\int^1_0 dx\ q^{val}(x,\mu_0^2) = 1 \;\;\; {\rm and}\;\;\;
\int^1_0 dx\ x [q^{val}(x,\mu_0^2) + \bar q^{val}(x,\mu_0^2)] 
= 1,
\label{MomSR}\end{equation}
where $\bar q^{val}(x)$ is the antiquark distribution. 
The fact that, 
at the low momentum scale $\mu_0$, 
the whole momentum in the pion is carried off by the valence
quarks is due to the quenched approximation used, 
when only valence quark-antiquark intermediate states are
included and all intrinsic quark-gluon sea states are neglected.

Thus, in the quenched approximation the dynamical information
contained in the first two moments is strongly restricted by the
symmetries and kinematics, and as a result, it is model independent.
The nontrivial dynamics is contained in the moments with $N> 2$.
The general structure of the moments of the structure function (SF),
from eq.~(\ref{MomIns}),
can be written in the form
\begin{equation} \displaystyle
A_N^{SF}(\mu_0^2) =  \frac{1}{2^{N-1} } \sum_{i=0}^{[\frac{N-1}{2}]}
\frac{1}{2i+1} {N-1 \choose 2i} J_i^{SF}(\mu_0^2),\ \ \ \ \
N=1,2,...
\label{AnSFexp}\end{equation}
with the coefficients $J_i^{SF}$ given by
\begin{equation}
J_i^{SF}(\mu_0^2) = \frac{1}{I_{g\pi}(0)}
\left\{ \int\limits_{0}^{\infty} \frac{dk\ k^{4i+3}
\tilde Q(k)}{(k^2+m_q^2)^{2i+3} }
\Big[2k^2 + (2i+3)m_q^2 \Big]  +...\right\},
\label{JiSF}\end{equation}
where
the vertex terms with derivatives, like that
appearing in eq.~(\ref{I2mpi=0}) for $I_{g\pi}(0)$,
are denoted by dots. In eq.~(\ref{AnSFexp}), the square brackets
$[...]$ mean the integer part of the number, and $\displaystyle {a
\choose b}$ are the binomial coefficients.

It is instructive to consider two extreme cases, depending on the
physics under consideration.  If the QCD vacuum were a very dense
medium, $\eta >> 1$, then $J_i^{SF} = 0$ for all $i$ except $i=0$.
As a result, it leads to the set of moments $A_N = 1/2^{(N-1)}$ for
all $n$ and to a quark distribution which has the form of a delta
function:  $q(x) = \delta(x-1/2)$.  This extreme case corresponds to
the heavy quark limit, and the coefficients $J_i^{SF}$ represent
consequent corrections in inverse powers of the heavy quark mass:
$\displaystyle \sim (\langle k^2\rangle/m_q^2)^i$.  In the opposite
extreme case of a very dilute vacuum $\eta << 1$ one gets $J_i^{SF}
= 1$ for all $i$ and $A_N = 1/N$ for the moments. This extreme case
corresponds to the momentum independent quark mass and provides
flat quark distribution $q(x) = 1$.  Moreover, the first term
in eq.~(\ref{JiSF}) dominates over the terms indicated by dots,
since the latter are small of an order of $O(\rho_c m_q)$.  A
realistic situation seems to be somewhere in-between these two
extremes.  Note that the role of pion mass is negligible, but the
interplay of the effective quark mass and the slope of the
nonlocality in $\tilde Q(k)$ has an important effect.

\section{Quark distribution function and its QCD evolution}

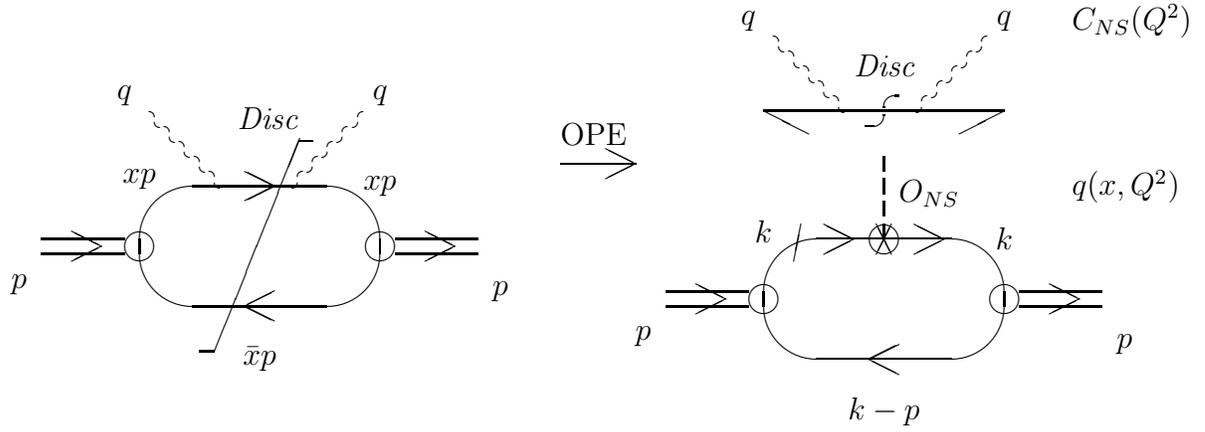
\begin{figure}
\begin{minipage}{4in}
\begin{center}
\unitlength=1.00mm
\special{em:linewidth 0.6pt}
\linethickness{0.6pt}
\begin{picture}(149.00,87.00)
\put(117.00,50.00){\oval(32.00,16.00)[]}
\put(133.00,50.00){\circle{4.00}}
\put(101.00,50.00){\circle{4.00}}
\put(117.00,58.00){\circle{4.00}}
\put(99.00,51.00){\line(-1,0){11.00}}
\put(88.00,51.00){\line(0,0){0.00}}
\put(99.00,49.00){\line(-1,0){11.00}}
\put(92.00,52.00){\line(2,-1){4.00}}
\put(96.00,50.00){\line(-2,-1){4.00}}
\put(146.00,51.00){\line(-1,0){11.00}}
\put(135.00,51.00){\line(0,0){0.00}}
\put(146.00,49.00){\line(-1,0){11.00}}
\put(139.00,52.00){\line(2,-1){4.00}}
\put(143.00,50.00){\line(-2,-1){4.00}}
\put(116.00,60.00){\line(1,-2){2.00}}
\put(116.00,56.00){\line(1,2){2.00}}
\put(85.00,45.00){\makebox(0,0)[cc]{$p$}}
\put(149.00,44.00){\makebox(0,0)[cc]{$p$}}
\put(101.00,59.00){\makebox(0,0)[cc]{$k$}}
\put(133.00,58.00){\makebox(0,0)[cc]{$k$}}
\put(117.00,35.00){\makebox(0,0)[cc]{$k-p$}}
\put(109.00,60.00){\line(2,-1){4.00}}
\put(113.00,58.00){\line(-2,-1){4.00}}
\put(121.00,60.00){\line(2,-1){4.00}}
\put(125.00,58.00){\line(-2,-1){4.00}}
\put(119.00,44.00){\line(-2,-1){4.00}}
\put(115.00,42.00){\line(2,-1){4.00}}
\put(117.00,58.00){\line(0,1){2.00}}
\put(117.00,61.00){\line(0,1){2.00}}
\put(117.00,64.00){\line(0,1){2.00}}
\put(117.00,67.00){\line(0,1){2.00}}
\put(28.50,65.50){\oval(1.00,1.00)[lb]}
\put(27.50,66.50){\oval(1.00,1.00)[rt]}
\put(26.50,67.50){\oval(1.00,1.00)[lb]}
\put(25.50,68.50){\oval(1.00,1.00)[rt]}
\put(24.50,69.50){\oval(1.00,1.00)[lb]}
\put(23.50,70.50){\oval(1.00,1.00)[rt]}
\put(22.50,71.50){\oval(1.00,1.00)[lb]}
\put(21.50,72.50){\oval(1.00,1.00)[rt]}
\put(20.50,73.50){\oval(1.00,1.00)[lb]}
\put(19.50,74.50){\oval(1.00,1.00)[rt]}
\put(38.50,65.50){\oval(1.00,1.00)[rb]}
\put(39.50,66.50){\oval(1.00,1.00)[lt]}
\put(40.50,67.50){\oval(1.00,1.00)[rb]}
\put(41.50,68.50){\oval(1.00,1.00)[lt]}
\put(42.50,69.50){\oval(1.00,1.00)[rb]}
\put(43.50,70.50){\oval(1.00,1.00)[lt]}
\put(44.50,71.50){\oval(1.00,1.00)[rb]}
\put(45.50,72.50){\oval(1.00,1.00)[lt]}
\put(46.50,73.50){\oval(1.00,1.00)[rb]}
\put(47.50,74.50){\oval(1.00,1.00)[lt]}
\put(34.00,57.00){\oval(32.00,16.00)[]}
\put(50.00,57.00){\circle{4.00}}
\put(18.00,57.00){\circle{4.00}}
\put(16.00,58.00){\line(-1,0){11.00}}
\put(5.00,58.00){\line(0,0){0.00}}
\put(16.00,56.00){\line(-1,0){11.00}}
\put(9.00,59.00){\line(2,-1){4.00}}
\put(13.00,57.00){\line(-2,-1){4.00}}
\put(63.00,58.00){\line(-1,0){11.00}}
\put(52.00,58.00){\line(0,0){0.00}}
\put(63.00,56.00){\line(-1,0){11.00}}
\put(56.00,59.00){\line(2,-1){4.00}}
\put(60.00,57.00){\line(-2,-1){4.00}}
\put(2.00,52.00){\makebox(0,0)[cc]{$p$}}
\put(66.00,51.00){\makebox(0,0)[cc]{$p$}}
\put(18.00,66.00){\makebox(0,0)[cc]{$xp$}}
\put(50.00,65.00){\makebox(0,0)[cc]{$xp$}}
\put(34.00,42.00){\makebox(0,0)[cc]{$\bar xp$}}
\put(32.00,67.00){\line(2,-1){4.00}}
\put(36.00,65.00){\line(-2,-1){4.00}}
\put(36.00,51.00){\line(-2,-1){4.00}}
\put(32.00,49.00){\line(2,-1){4.00}}
\put(107.00,72.00){\line(-2,1){6.00}}
\put(101.00,75.00){\line(1,0){32.00}}
\put(133.00,75.00){\line(-2,-1){6.00}}
\put(111.50,75.50){\oval(1.00,1.00)[lb]}
\put(110.50,76.50){\oval(1.00,1.00)[rt]}
\put(109.50,77.50){\oval(1.00,1.00)[lb]}
\put(108.50,78.50){\oval(1.00,1.00)[rt]}
\put(107.50,79.50){\oval(1.00,1.00)[lb]}
\put(106.50,80.50){\oval(1.00,1.00)[rt]}
\put(105.50,81.50){\oval(1.00,1.00)[lb]}
\put(104.50,82.50){\oval(1.00,1.00)[rt]}
\put(103.50,83.50){\oval(1.00,1.00)[lb]}
\put(102.50,84.50){\oval(1.00,1.00)[rt]}
\put(121.50,75.50){\oval(1.00,1.00)[rb]}
\put(122.50,76.50){\oval(1.00,1.00)[lt]}
\put(123.50,77.50){\oval(1.00,1.00)[rb]}
\put(124.50,78.50){\oval(1.00,1.00)[lt]}
\put(125.50,79.50){\oval(1.00,1.00)[rb]}
\put(126.50,80.50){\oval(1.00,1.00)[lt]}
\put(127.50,81.50){\oval(1.00,1.00)[rb]}
\put(128.50,82.50){\oval(1.00,1.00)[lt]}
\put(129.50,83.50){\oval(1.00,1.00)[rb]}
\put(130.50,84.50){\oval(1.00,1.00)[lt]}
\put(114.50,74.50){\oval(5.00,3.00)[rb]}
\put(119.00,75.50){\oval(4.00,3.00)[lt]}
\put(16.00,77.00){\makebox(0,0)[cc]{$q$}}
\put(50.00,77.00){\makebox(0,0)[cc]{$q$}}
\put(99.00,87.00){\makebox(0,0)[cc]{$q$}}
\put(133.00,87.00){\makebox(0,0)[cc]{$q$}}
\put(74.00,68.00){\line(1,0){10.00}}
\put(80.00,70.00){\line(2,-1){4.00}}
\put(84.00,68.00){\line(-2,-1){4.00}}
\put(74.00,72.00){\makebox(0,0)[lc]{OPE}}
\put(142.00,87.00){\makebox(0,0)[lc]{$C_{NS}(Q^2)$}}
\put(142.00,65.00){\makebox(0,0)[lc]{$q(x,Q^2)$}}
\put(105.00,55.00){\line(1,5){1.00}}
\put(119.00,64.00){\makebox(0,0)[lc]{$O_{NS}$}}
\put(28.00,43.00){\line(2,5){11.33}}
\put(39.33,71.00){\line(1,0){1.67}}
\put(28.00,43.00){\line(-1,0){2.00}}
\put(35.00,74.00){\makebox(0,0)[cc]{{\it Disc}}}
\put(117.00,81.00){\makebox(0,0)[cc]{{\it Disc}}}
\end{picture}
\end{center}
\end{minipage}
 \caption[dummy0]{
 Graphical representation of the operator product expansion. The
left hand side of this diagram is the imaginary part ({\it
Disc}ontinuity) of the forward scattering amplitude. Within OPE it
is represented by the convolution of the Wilson coefficient function
$C_{NS}(Q^2)$ of a ``hard" parton subprocess (upper block of the
right diagram) and the ``soft" parton distribution function
$q(x,Q^2)$ (lower block of the right diagram). The constituent quark
and  pion are depicted by solid and double lines, respectively. The
wavy line denotes the virtual photons. $O_{NS}$ is the local
operator and cross on the quark line correspond to $\delta(x-(kn))$.

  \label{fig:OPE}}
\end{figure}

 Let us now turn our attention to the quark distribution itself.
This distribution for the pion with 4-momentum $p$ is given by
(see a graphical representation in Fig. \ref{fig:OPE})
\begin{eqnarray}  \displaystyle
&&q(x;\mu_0^2) p^\mu  = 2N_c g^2_{\pi q\bar q}
\int \frac{d^4k}{(2\pi)^4} \delta\big[x - 1 - (k\cdot n)\big] \cdot
\label{qx} \\
&&\cdot Tr\left\{\gamma_5 S(k+p)
\left[ \tilde F^2(k,k+p;\mu_0^2) \gamma_\mu S(k+p)
- 2 \left(\frac{\partial \tilde {F^2}(k,k+p;\mu_0^2)}
{\partial(k+p)^2}\right)
(k+ p)_\mu \right] \gamma_5 S(k) \right\},
\nonumber\end{eqnarray}
where $q(x) = \bar u(x)_{val} = d(x)_{val}$ for $\pi^-$.  Here, we
arrive at the quark distribution defined in a similar manner as that
used in \cite{Baul80,MikhRad85}.  The $\delta\big[x - 1 - (k\cdot
n)\big] -$ function appearing in eq.~(\ref{qx}) represents the
effective vertex related to the composite operator $O_N^{j} $ given
by eq.~(\ref{OPE}). It accumulates information about all moments of
the distribution function (\ref{MomIns})  and is related to them by a
Mellin transformation if the light\--\-cone vector $n$ is normalized
by $(pn) =1$.
The light-cone vector $n$ serves to project out in Eqs.
(\ref{MomIns}) and (\ref{qx}) symmetric traceless tensors. It can be
easily shown that the first moments of $q(x)$ will reproduce the
parton sum rules eq.~(\ref{A1-2}).

To calculate the $k-$integral in eq.~(\ref{qx}), we use
$\alpha-$representation for the propagators\footnote{
For details, see ref.~\cite{MikhRad85}.}, 
\begin{equation}
\frac{1}{k^2+m^2} = \int\limits_{0}^{\infty} d\alpha\ 
\exp{\left[-\alpha (k^2+m^2)\right]},
\label{alpha}\end{equation}
and for the vertex $\delta-$function,
\begin{equation}
\delta\big[x - (k\cdot n)\big] = \frac{1}{2\pi }
\int\limits_{-\infty}^{\infty} d\alpha\ \exp{[i \alpha(x-k\cdot n)]}.
\label{aldel}\end{equation}
Then, a direct calculation from eq.~(\ref{qx}) provides the result
for the quark distribution, which in the massless case
($m_\pi^2 = -p^2 = 0$) is reduced to
\begin{eqnarray}  \displaystyle
 q(x,\mu_0^2) & = &
\frac{ N_c g^2_{\pi q\bar q} }{2 \pi^2}
\int\limits_{0}^{\infty} \int\limits_{0}^{\infty}
d\nu_1 d\nu_2\
F(\nu_1) F(\nu_2) \exp\Big( \frac{m_q^2}{\nu_1} + 
\frac{m_q^2}{\nu_2}\Big)
\cdot
\label{qx1}\\
&&\cdot \left\{ \left[ E_1\left( \frac{m_q^2}{x\nu_1}\right)   +
\bar x \exp\left(-\frac{m_q^2}{x \nu_1} \right)\right]
\Theta(\bar x \nu_2 \geq x \nu_1) +
(x \leftrightarrow \bar x)\right\}.
\nonumber\end{eqnarray}
In the above equation, $\bar x = 1-x$, $E_1(z)$ is the integral
exponential, and $F(\nu)$ is the correlation function related to
the vertex function $\tilde Q(k) $ by the Laplace transformation. 
The vertex $\tilde Q(p)$, in the essential region of $p$
($0\le p\le 4/\rho_c$)  is approximated by
\begin{equation}
\tilde Q(p) = 4.5 \exp{(-1.9\rho_c p)}
- 3.5 \exp{(-3.6\rho_c p)},
\end{equation}
which leads to the Laplace transform, 
\begin{equation}
F(\nu) = \frac{\rho_c}{\sqrt{\pi}\sqrt{\nu}}
\left[8.55  \exp(- 0.9\rho_c^2 \nu)
-  12.6 \exp(- 3.24\rho_c^2 \nu)\right].
\label{Fsf}\end{equation}
Let us stress that the expressions eq.~(\ref{AnSFexp}) for the
moments and eq.~(\ref{qx1}) for the valence-quark distribution
in the pion are universal ones and valid for any shape of the
functions $\tilde Q(k)$ and $F(\nu)$, which in turn are determined
by the concrete model of the quark-pion dynamics.

\begin{figure}
\vspace*{0.5cm}
\epsfxsize=15cm
\epsfysize=15cm
\centerline{\epsfbox{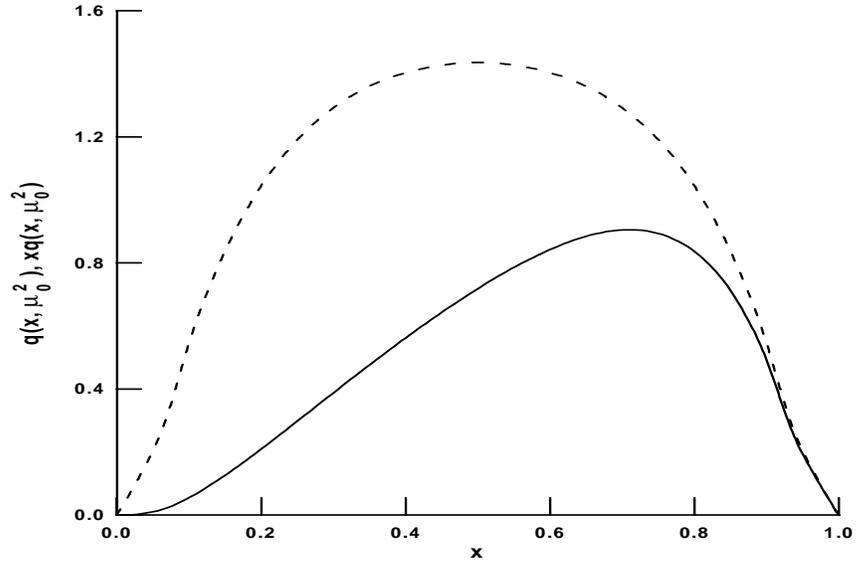}}
 \caption[dummy0]{
The valence-quark distribution function (DF) $q(x;\mu_0^2)$ (dashed
line) and the quark momentum distribution function (MDF)
$xq(x;\mu_0^2)$ (solid line) for the pion as a function of the
longitudinal momentum fraction $x$ at the low momentum scale
$\mu^2_0 = 0.19$ GeV$^2$  and
density parameter ${\rho}_c m_q = 0.39$.

\label{fig:SFPI} }
\end{figure}

The quark distribution $q(x,\mu _0^2)$ and the momentum distribution
(structure function) $x q(x,\mu _0^2)$ are shown graphically in
Fig.~\ref{fig:SFPI}. We have to note that the shape of the
distribution is quite stable with respect to changes of the
instanton model parameters, if they are fixed to reproduce the pion
low energy properties.  We remind that we are computing only the
leading-twist distributions at a low normalization point $\mu_0 \sim
\rho_c^{-1}$ rather than the full structure functions which contain
also higher-twist corrections. The latter may be large at low $q^2$.
We have also to note that these results differ strongly from those
obtained in calculations with   the NJL model \cite{DRA-KS96} which
yield  distributions that are rather consistent with the strict
chiral limit $q(x,\mu_0^2) \approx 1$.

The computed distributions are then used as initial conditions for
the perturbative evolution to higher values of $Q^2$, where the
power corrections are expected to be suppressed, so that one can
compare them with the available experimental data.  Actually, we
compare our theoretical predictions with the phenomenological
analysis by Sutton {\it et al} \cite{Sutt92} of the data taken from
Drell - Yan and prompt photon experiments performed by the groups
NA10 (CERN) and E615 (Fermilab) \cite{PiSFex}.

\begin{figure}
\vspace*{0.5cm}
\epsfxsize=15cm
\epsfysize=15cm
\centerline{\epsfbox{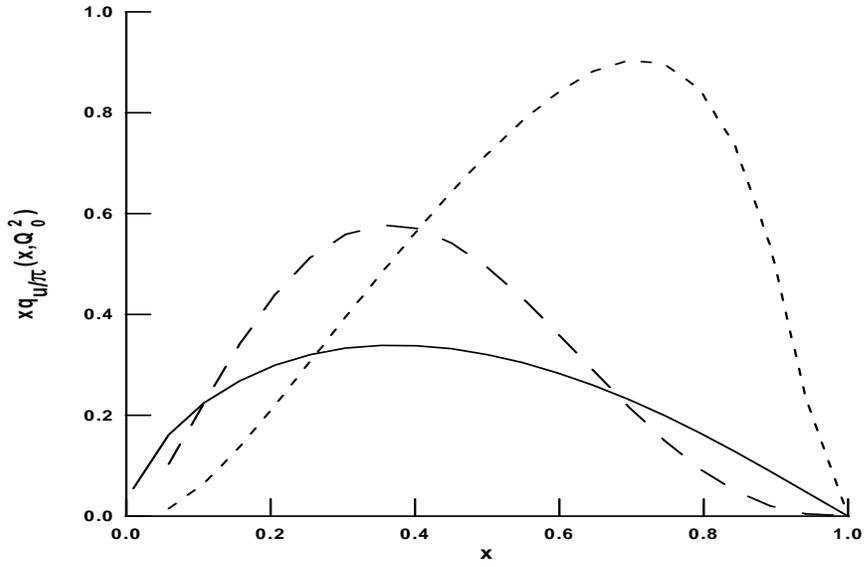}}
 \caption[dummy0]{
The quark momentum distribution function $ xq(x;Q_0^2)$ (long
dashed line) for the pion as a function of the variable $x$ evolved
to the momentum scale $Q^2_0 = 4$ GeV$^{2}$ (LO approximation),
using ${\rho}_c m_q = 0.39$ for the density parameter.  The solid
line denotes the phenomenological curve \cite{Sutt92} on the same
scale $Q_0^2$, extracted from the data. The short - dashed line shows the
same distribution on the low momentum scale $\mu^2_0 = 0.19$
GeV$^2$.

  \label{fig:SFPIev} }
\end{figure}

The form of the evolved distribution $q(x,Q_0^2)$ at the momentum
scale $Q_0^2 = 4$ GeV$^2$ is reconstructed from its moments evolved
to this scale in the leading order (LO) and next\--\-to -leading
order (NLO) perturbative QCD in the $\overline{MS}$ scheme by using
the first six Jacobi polynomials.  To this goal we use the
well\--\-known expressions \cite{YndBook} for the perturbatively
calculable coefficient function of the process $\displaystyle C_i^N
= C_{0i}^N + \frac{\alpha _{s}(Q^{2})}{4\pi}C_{1i}^N$ and the
anomalous dimensions $\gamma_{(n)}$ calculated up to LO and NLO.
Thus, the final result for the moments obtained from the
factorization procedure is
\begin{equation}
A_N(Q^2) = \sum\limits_{i}C_i^N(Q^2,\mu^2) O_i^N(\mu^2)=
 \int\limits_{0}^{1} dx\ x^N q(x,Q^2).
\label{FP}\end{equation}
In performing the evolution analysis we choose a low momentum scale
$\displaystyle \mu_0^2 = 0.19 \pm 0.05\ GeV^2$, and a set for the
QCD scale parameter $\Lambda_{QCD}^{(3)} = 0.19\ GeV$ in order to be
consistent with \cite{Sutt92}.  The resulting distribution
$q(x,Q_0^2)$ is shown in Fig. \ref{fig:SFPIev} together with the
phenomenological curve derived from the data in \cite{Sutt92}. The
initial distribution function at the low-momentum scale $\mu _0^2$
is also shown for comparison.

The values of the first moments of the pion quark distribution at
$Q^2_0 = 4\ GeV^2$ calculated in LO and NLO are shown in Table 2.
The error bars quoted in Table 2 for our calculations are due to
accepted uncertainty in the choice of the initial scale of evolution
$\mu_0$.  These values should be compared with those obtained from
the phenomenological analysis \cite{Sutt92} and from LQCD
simulations \cite{PiRhoLat}.  In Table 2 we also include the moments
of  quark distribution in the pion obtained from the parametrization
\cite{ReyaPi}.

\vskip 0.5cm

\begin{center}
 {\bf Table 2.}
 The values of the first moments at $Q_0^2 = 4$ GeV$^2$.\\[0.5cm]
\begin{tabular}
{||l          |c       |c       |c    |c      |c||}     \hline \hline
      &              &              &
&                    &                                \\
      & LO           & NLO         &LQCD\cite{PiRhoLat}
& Exp. fit\cite{Sutt92}& Exp. fit\cite{ReyaPi}            \\
      & (this        & calculations)&
&                    &                                \\ \hline
&                    &              &
&                    &                                \\
$A_2(Q^2_0) $&$ 0.321\pm0.02 $&$ 0.293\pm0.03 $&$ 0.279\pm 0.083 $
&$ 0.230\pm 0.01 $   &$ 0.193\pm 0.01 $                \\
&                    &              &
&                    &                                 \\
$A_3(Q^2_0) $&$ 0.150\pm0.02 $&$ 0.112\pm0.025 $&$ 0.107\pm 0.035 $
&$ 0.101\pm 0.005$   &$ 0.083\pm 0.005$                \\
&                    &                    &
&                    &                                 \\
$A_4(Q^2_0) $&$ 0.083\pm0.015 $&$ 0.057\pm0.025 $&$ 0.048\pm 0.020$
&$ 0.057\pm 0.005$   &$ 0.046\pm 0.005$                \\ \hline
	\end{tabular}
\end{center}
\vskip 0.5cm

Let us finally discuss the uncertainties of the QCD evolution from
the low momentum scale $\mu_0$.  As we see from Table 2, the
difference of the LO and NLO results is in the range of $30\%$. It
turns out that the use of a larger initial evolution scale, say $\mu
_{0}^{2} \geq 0.3$~GeV${}^{2}$, gives a rather good convergence with
deviations less than $10\%$,  whereas in the opposite case, i.e.,
for scales smaller than about $0.1$~GeV${}^{2}$ the deviations
increase and perturbative evolution loses any sense. This behavior
has also been observed in analyzes within the NJL model
\cite{DRA-KS96}.

The comparison shows that our calculations, in particular in NLO,
are consistent with the phenomenological analysis of \cite{Sutt92}
and fairly close to the LQCD results. Both theoretical approaches
(LQCD and the instanton model) predict moment values systematically
larger than the phenomenological one.  One of the reasons for this
disagreement may be traced to the quenched approximation which does
not take into account any sea-quark contributions at the initial
scale, attributing in this way the whole pion momentum to the
valence quarks. Indeed, the origin of the $A_{2}$ moment at the
initial scale (in the quenched approximation) and its subsequent
evolution is purely kinematic and does not depend on the details of
the model.  In principle, one could match the valence momentum
fraction derived in our calculation with that determined in
\cite{Sutt92} by shifting the initial value $\mu _{0}^{2}$ down to
$0.01$~GeV${}^{2}$ (see, for instance, \cite{DRA-KS96}a). However,
to start a perturbative evolution from this very low scale is
formally incorrect and technically amounts to a rather unstable
procedure.  In our opinion, it is more realistic to expect that
including in our analysis contributions from the sea, which may have
momentum fractions of an order of $10\%$, the agreement between the
theoretical predictions and the phenomenological analysis
can be considerably improved.  In addition, the effect of
nonperturbative evolution~\cite{Genov96} from an initial scale $\mu
_{0}^{2}$ up to $Q^{2} \sim 1$~GeV${}^{2}$ could be important.

\section{Results and discussions}

 In summary, we have presented theoretical predictions for the
valence-quark distribution function, eq.~(\ref{qx1}), and its
moments, eq.~(\ref{AnSFexp}), for the pion. The calculations are
based on the instanton model of the QCD vacuum as a candidate
treatment of nonperturbative dynamics, expressing the observable
hadron properties in terms of fundamental characteristics of the
vacuum state. We found that the instanton model describes well
the vacuum expectation values of the lowest-dimension
quark-gluon operators and the pion low-energy observables.  
To obtain these results, we have used gauge invariant forms for
the dynamically generated quark masses and quark pion vertex.
Thus, we are led to express the form of the pion quark
distribution function in terms of the effective instanton size
$\rho _{c}$, and the quark-mass parameter $m_q$.  
The pion quark distribution function extracted corresponds
to a low normalization scale, where the effective instanton
approach is justified.  It is shown that the validity of parton
sum rules for the isospin and total momentum distribution is a
consequence of the compositeness condition and the strict
implementation of gauge invariance.  We have used techniques to
derive these results which constitute a complementary approach
to lattice simulations and to phenomenological fits to
experimental data.  Using this distribution function as an
input, we obtained the quark distribution function in
the pion via standard perturbative evolution to higher momentum
values, accessible by experiment.  A reasonable agreement with
the data was found.  In fact, the calculations are performed in
the quenched approximation, where the effect of intrinsic
quark-gluon sea is neglected. We expect that the effects of 
the intrinsic quark component of the pion wave function and the
nonperturbative evolution at intermediate energy scale provide 
a better agreement between theoretical predictions and 
phenomenological analysis.

\vspace{1cm}

\centerline{\bf Acknowledgments}
\vspace{2mm}
The authors are grateful to I.V. Anikin, M. Birse, D.I. Diakonov,
S.B. Gerasimov, P. Kroll, A.E. Maximov, S.V. Mikhailov, M. Polyakov,
R. Ruskov, N.G. Stefanis, M.K. Volkov for fruitful discussions of
the results.  One of us (A.E.D.) thanks the members of the particle
physics groups of the Wuppertal University and Instituto de F\'\i
sica Te\'orica, UNESP, (S\~ao Paulo) for their hospitality and
interest in the work.  This investigation (A.E.D.) was supported in
part by the Russian Foundation for Fundamental Research (RFFR)
96-02-18096  and 96-02-18097, St. - Petersburg center for
fundamental research grant: 95-0-6.3-20 and
by the Heisenberg-Landau program.
L.T. also thanks
partial support received from Funda\c c\~ao de Amparo \`a Pesquisa
do Estado de S\~ao Paulo (FAPESP) and from Conselho Nacional de
Desenvolvimento Cient\'\i fico e Tecnol\'ogico do Brasil.

\newpage


\vskip 0.5cm

\end{document}